# Mutual orientation of the Poynting vector and the group velocity vector of electromagnetic waves in bigyrotropic medium

**Edwin H. Lock and Sergey V. Gerus**

(*Kotel'nikov Instutute of Radio Engineering and Electronics (Fryazino branch), Russian Academy of Sciences, Fryazino, Moscow region, Russia*)

The collinearity of the Poynting vector **P** and the group velocity vector **U** of electromagnetic waves in a bihyrotropic medium characterized by second rank Hermitian tensors of dielectric and magnetic permittivities is theoretically proved. It is shown that these vectors will be collinear in any anisotropic medium having either Hermitian or symmetric dielectric and magnetic permittivities tensors.

## 1. Introduction

Electromagnetic waves propagating in ferrites, antiferromagnetics, plasmas and in various structures based on these media have been studied since the middle of the last century (see, for example, monographs [1 - 4] and their reference list). The characteristics of electromagnetic waves in these gyrotropic media are still the subject of many studies, mainly focused on the design of various microwave devices (see, for example, reviews [5 - 7]). Obviously, a theoretical consideration of electromagnetic waves properties in bihyrotropic medium, special cases of which are the listed gyrotropic media, would help to develop these studies, because the discovered new properties of electromagnetic waves would occur simultaneously in all these media. In addition, both the Maxwell's equations themselves for a bihyrotropic medium and the solutions obtained from them are most mathematically symmetric, making it easy to write down solutions for any gyrotropic medium as a result.

Propagation of electromagnetic waves in unbounded bihyrotropic medium has been studied earlier in works [1, 8 – 10]. In particular, it has been shown that Maxwell's equations for bihyrotropic medium one can bring to a system of two second order differential equations, which contain only electric and magnetic microwave field components parallel to the constant homogeneous magnetic field vector $\mathbf{H_0}$ [1]. The dispersion equation for electromagnetic waves propagating in bihyrotropic medium was presented in the monograph [8], and then on the basis of this equation the isofrequency surfaces and dependences for various types of electromagnetic waves (including spin waves), propagating in an unbounded ferromagnetic space (which is a special case of a bihyrotropic medium) were calculated and it was shown that spin waves are characterized by unclosed isofrequency dependences and the presence of frequency-dependent cutoff angles in the certain frequency interval [9]. Formulas for the Poynting vector $\mathbf{P}$, group velocity vector $\mathbf{U}$, volume energy density $w$ and for all components of electromagnetic wave propagating in an arbitrary direction in unbounded bihyrotropic medium were obtained in [10] and it was proved that corresponding components of the vectors $\mathbf{P}$ and $\mathbf{U}$ are proportional. In addition, the problem of electromagnetic wave propagation in an arbitrary direction along a tangentially magnetized bihyrotropic layer was solved analytically and it was found that the wave distribution over the layer thickness is described by two different wave numbers, which can take only real or only imaginary values [11].

Below the mutual orientation of the Poynting and the group velocity vectors of electromagnetic waves propagated in bihyrotropic medium is investigated on the basis of Maxwell's equations by means of the method used earlier in [12]. As will be shown, this method based on mathematical operations with vectors in Maxwell's equations allows one to prove relatively easily the collinearity of Poynting and group velocity vectors of electromagnetic waves in unbounded bihyrotropic medium independently of its specific parameters.

## 2. Proof of collinearity of Poynting vector P and group velocity vector U of electromagnetic waves in unbounded bihyrotropic medium.

Let a monochromatic electromagnetic wave with frequency $\omega$ = const, varying in time according to the harmonic law $\exp(i\omega t)$, propagate in an unbounded bihyrotropic medium that has a dedicated direction along the z-axis. It is known that such medium is characterized by dielectric and magnetic permittivities described by second rank Hermitian tensors:

$$\ddot{\mu} = \begin{vmatrix} \mu & i\nu & 0 \\ -i\nu & \mu & 0 \\ 0 & 0 & \mu_{zz} \end{vmatrix}, \tag{1}$$

$$\ddot{\varepsilon} = \begin{vmatrix} \varepsilon & ig & 0 \\ -ig & \varepsilon & 0 \\ 0 & 0 & \varepsilon_{zz} \end{vmatrix}. \tag{2}$$

An electromagnetic wave propagating in this medium should satisfy a system of Maxwell's equations for complex amplitudes

$$\begin{cases} \operatorname{rot} \mathbf{E} + i\omega \mathbf{B} / c = 0 \\ \operatorname{div} \mathbf{B} = 0 \\ \operatorname{rot} \mathbf{H} - i\omega \mathbf{D} / c = 0 \\ \operatorname{div} \mathbf{D} = 0 \end{cases}, \tag{3}$$

where $c$ is the speed of light in vacuum; $\mathbf{E}$ and $\mathbf{H}$ are the amplitudes of intensity vectors of microwave electric and magnetic field; $\mathbf{D}$ and $\mathbf{B}$ are the amplitudes of vectors of microwave electric and magnetic induction, which are related with $\mathbf{E}$ and $\mathbf{H}$ by the formulas

$$\mathbf{D} = \ddot{\varepsilon} \mathbf{E}, \tag{4}$$

$$\mathbf{B} = \ddot{\mu} \mathbf{H}. \tag{5}$$

We will look for solutions of the equations system (3) in the form of homogeneous plane electromagnetic wave with wave vector $\mathbf{k}$, i.e., we will assume that all components of the fields $\mathbf{E}$ and $\mathbf{H}$ — $E_x$ $E_y$, $E_z$, $H_x$, $H_y$ and $H_z$ — vary in space (as well as in time) according to a harmonic law in accordance with the expressions

$$\mathbf{E} = \mathbf{E_0} \exp(-i\mathbf{kr}), \tag{6}$$

$$\mathbf{H} = \mathbf{H_0} \exp(-i\mathbf{kr}). \tag{7}$$

By substituting expressions (6) and (7) for vectors E and H (and similar expressions for vectors D and B) directly into the first and third equations of system (3) and performing the rot operation, we obtain[1]

$$\mathbf{B} = \left[\mathbf{kE}\right] / k_0, \tag{8}$$

$$\mathbf{D} = -\left[\mathbf{kH}\right] / k_0, \tag{9}$$

where the notation $k_0 = \omega/c$ is used.

Note now that since the ends of all wave vectors of the electromagnetic wave under study lie on its iso-frequency surface (or wave vectors surface), a vector equal to the difference of two very close lying wave vectors **k** and **k'** will connect two very close points of the iso-frequency surface. That is, when vector **k'** tends to vector **k**, the difference of these two wave vectors – differential vector $d\mathbf{k}$ – will in the limit lie in the plane which is tangent to the iso-frequency surface at the point corresponding to vector **k** (the point to which vector **k** is directed). Therefore, if we prove that the scalar product of the Poynting vector **P** on any arbitrarily chosen vector $d\mathbf{k}$ is zero, it will mean that the Poynting vector of electromagnetic wave is always perpendicular to its iso-frequency surface. Here the term "any arbitrarily chosen vector" means the following: whichever arbitrary vector $d\mathbf{k}$ we choose, this vector will always lie in a plane tangent to the iso-frequency surface at a given point. Therefore, if we prove perpendicularity of the Poynting vector **P** to any arbitrary vector $d\mathbf{k}$, it will mean that vector **P** is perpendicular to the given plane and, consequently, to the iso-frequency surface of the electromagnetic wave.

By differentiating equations (8) and (9) (which will result in the appearance of the value d**k** in these equations), we obtain

$$d\mathbf{B} = \left[\mathbf{k}d\mathbf{E}\right] / k_0 + \left[d\mathbf{kE}\right] / k_0, \tag{10}$$

---

[1] Below we will use square brackets to write the vector product of vectors, while we will not use round brackets to write the scalar product of vectors (to avoid confusion with the use of round brackets in conventional mathematical expressions). In addition, for simplicity in expressions (8), (9) and further, we omit the "0" indices introduced in (6) and (7) for the amplitudes of vectors **E**, **H**, **D** and **B**.

$$dD = -[\mathbf{k}d\mathbf{H}]/k_0 + [\mathbf{H}d\mathbf{k}]/k_0. \tag{11}$$

Now, to make the vector product $\mathbf{E}$ and $\mathbf{H}$ defining the Poynting vector $\mathbf{P}$ appear in relations (10) and (11), we multiply scalarly equation (10) with the complex-conjugate value $\mathbf{H}*$ and equation (11) with the complex-conjugate value $\mathbf{E}*$:

$$\mathbf{H}*d\mathbf{B} = \mathbf{H}*[\mathbf{k}d\mathbf{E}]/k_0 + \mathbf{H}*[d\mathbf{k}\mathbf{E}]/k_0, \tag{12}$$

$$\mathbf{E}*d\mathbf{D} = -\mathbf{E}*[\mathbf{k}d\mathbf{H}]/k_0 + \mathbf{E}*[\mathbf{H}d\mathbf{k}]/k_0. \tag{13}$$

Using the vector multiplication rules and taking into account expressions (8) and (9), equations (12) and (13) can be written as

$$\mathbf{H}*d\mathbf{B} = \mathbf{D}*d\mathbf{E} + [\mathbf{E}\mathbf{H}*]d\mathbf{k}/k_0, \tag{14}$$

$$\mathbf{E}*d\mathbf{D} = \mathbf{B}*d\mathbf{H} + [\mathbf{E}*\mathbf{H}]d\mathbf{k}/k_0. \tag{15}$$

Since the tensors of the dielectric and magnetic permittivity $\ddot{\varepsilon}$ and $\ddot{\mu}$ for the investigated bihyrotropic medium are Hermitian, the following relations are valid:

$$\mathbf{H}*d\mathbf{B} = \mathbf{H}*d\left(\ddot{\mu}\mathbf{H}\right) = \mathbf{B}*d\mathbf{H}, \tag{16}$$

$$\mathbf{E}*d\mathbf{D} = \mathbf{E}*d\left(\ddot{\varepsilon}\mathbf{E}\right) = \mathbf{D}*d\mathbf{E}. \tag{17}$$

Since Re[$\mathbf{E}\mathbf{H}*$] ≡ Re[$\mathbf{E}*\mathbf{H}$], adding equations (14) and (15) with taking into account (16) and (17), we obtain

$$[\mathbf{E}\mathbf{H}*]\,d\mathbf{k} = 0 \text{ or } \mathbf{P}\,d\mathbf{k} = 0. \tag{18}$$

Now, we have proved that the Poynting vector $\mathbf{P}$ and any arbitrarily chosen vector $d\mathbf{k}$ (lying in the plane tangent to the iso-frequency surface at the point corresponding to vector $\mathbf{k}$) are perpendicular or that the Poynting vector $\mathbf{P}$ is always perpendicular to the iso-frequency surface.

Since the group velocity vector $\mathbf{U}$ of the wave is determined by the expression[2] [10, 13]

---

[2] Using formula (19) for the group velocity vector $\mathbf{U}$, we assume that the concept of "group velocity" can be used for the considered bihyrotropic medium in accordance with the physical meaning of this concept in [10, 13].

$$\mathbf{U} = \frac{d\omega}{d\mathbf{k}} = \text{grad}_\mathbf{k}\,\omega = \frac{\partial\omega}{\partial k_x}\mathbf{x_0} + \frac{\partial\omega}{\partial k_y}\mathbf{y_0} + \frac{\partial\omega}{\partial k_z}\mathbf{z_0}, \qquad (19)$$

it is obvious that vector $\mathbf{U}$ is always perpendicular to the iso-frequency surface according to definition (as a gradient of the iso-frequency surface).

Thus, we have proved that vectors $\mathbf{P}$ and $\mathbf{U}$ are always collinear.

### 3. Conclusions

Based on mathematical operations with vectors in Maxwell's equations, the collinearity of the Poynting vector $\mathbf{P}$ and the group velocity vector $\mathbf{U}$ of electromagnetic waves in unbounded bihyrotropic medium is proved. Since the fulfillment of relations (16) and (17) provide Hermitian properties of the tensors and the bihyrotropic medium, all the above formulas and statements are valid for any anisotropic medium whose tensors of dielectric and magnetic permittivity are either Hermitian or symmetric.

It should be noted, however, that the collinearity of vectors P and U means that these vectors can be oriented either in the same direction or opposite to each other. It is impossible to prove that the vectors P and U in a bihyrotropic medium are always oriented in the same direction by means of above method. Moreover, the exact formulas for the Poynting vector P, the group velocity vector U and the volume energy density w (equal to the ratio of P and U) obtained in [10] also do not allow to prove that $\mathbf{P}$ and $\mathbf{U}$ are directed by the same way. In particular, the formula for the volume energy density w includes values which probably can be negative (see formulas (50) - (52) in [10]).

### Funding